\documentclass[conference]{IEEEtran}
\IEEEoverridecommandlockouts
\usepackage{cite}
\usepackage{amsmath,amssymb,amsfonts}
\usepackage{algorithmic}
\usepackage{graphicx}
\usepackage{textcomp}
\usepackage{subfigure}
\usepackage{url}
\usepackage{xcolor}
\usepackage{stfloats}
\usepackage{multirow}
\usepackage{comment}
\usepackage{amssymb}
\usepackage{pifont}
\usepackage{tabularx}
\newcommand{\xmark}{\ding{55}}
\newcommand{\cmark}{\ding{51}}

\def\BibTeX{{\rm B\kern-.05em{\sc i\kern-.025em b}\kern-.08em
    T\kern-.1667em\lower.7ex\hbox{E}\kern-.125emX}}

\usepackage{multirow}
\usepackage{makecell}
\usepackage{hyperref}

\begin{document}
\title{DSSCNet: A Transfer Learning Framework for Cross-Corpus Dysarthric Speech Severity Classification}

\author{
    \IEEEauthorblockN{Arnab Kumar Roy\IEEEauthorrefmark{1}, Hemant Kumar Kathania\IEEEauthorrefmark{2},
    Paban Sapkota\IEEEauthorrefmark{2},
    Sudarsana Reddy Kadiri\IEEEauthorrefmark{3}, and Shrikanth Narayanan\IEEEauthorrefmark{3}}
    \\
    \IEEEauthorblockA{\IEEEauthorrefmark{1} Department of Computer Science and Engineering, Sikkim Manipal Institute of Technology, India.\\
Email: arnab\_202000152@smit.smu.edu.in}
    \IEEEauthorblockA{\IEEEauthorrefmark{2} Department of Electronics and Communication Engineering, National Institute of Technology Sikkim, India.  \\
    Emails: phec230006@nitsikkim.ac.in, hemant.ece@nitsikkim.ac.in}

    \IEEEauthorblockA{\IEEEauthorrefmark{3} Signal Analysis and Interpretation Laboratory (SAIL), University of Southern California, Los Angeles, USA.\\
    Emails: skadiri@usc.edu, shri@usc.edu}
}

\maketitle

\begin{abstract}

Dysarthric speech severity classification is challenging due to speaker variability, class imbalance, and limited datasets. This study introduces DSSCNet, a deep learning model that employs transfer learning and multi-corpus learning to enhance speaker-independent classification. By pre-training on one dysarthric speech corpus and fine-tuning on another, DSSCNet achieves improved feature extraction and cross-corpus generalization. Experimental results demonstrate that DSSCNet outperforms state-of-the-art models for speaker-independent severity classification, achieving 75.80\% accuracy on TORGO and 68.25\% on UA-Speech, significantly reducing misclassification errors. The findings confirm that leveraging knowledge transfer between datasets improves model robustness, making DSSCNet well-suited for automated dysarthria assessment. This research contributes to the development of more effective assistive speech technologies for individuals with speech impairments.

\end{abstract}

\begin{IEEEkeywords}
dysarthria, severity classification, transfer-learning, fine-tuning, convolutional neural network, squeeze-excitation, residual network.
\end{IEEEkeywords}
\section{Introduction}
\label{sec:introduction}
Dysarthria is a neuromotor speech disorder caused by impaired muscle control, affecting articulation, phonation, and overall speech intelligibility \cite{Dysarthr64:online}. 
The severity of dysarthria varies across individuals, ranging from mild to severe, depending on the extent of speech degradation. 
Accurate severity classification is essential in both clinical settings and assistive technologies, aiding speech-language pathologists in developing personalized therapy \cite{stipancic2021you, jayaraman2023dysarthria} plans and enhancing Automatic Speech Recognition (ASR) systems for dysarthric speakers \cite{yeo2023automatic}. A reliable classification framework enables adaptive speech processing, improving accessibility for individuals with speech impairments while advancing the development of robust human-computer interaction systems.



Recent advancements in dysarthric speech severity classification have been largely driven by progress in deep learning and speech processing methodologies. Traditional machine learning approaches primarily relied on handcrafted acoustic features such as Mel-Frequency Cepstral Coefficients (MFCCs), Linear Predictive Coding (LPC), and prosodic features, which were typically processed using classifiers like Support Vector Machines (SVMs) \cite{hearst1998support}, Random Forests (RF) \cite{breiman2001random}, and Gaussian Mixture Models (GMMs) \cite{reynolds2009gaussian}. While these methods showed reasonable performance, they were limited by speaker variability, noise sensitivity, and poor generalization across severity levels. To overcome these challenges, deep learning techniques have increasingly been employed. Convolutional Neural Networks (CNNs) have been used to extract hierarchical spectral features, as demonstrated in \cite{shih2022dysarthria}, where a CNN-GRU model effectively captured dysarthric speech characteristics. More recent approaches incorporate multi-head attention and multi-task learning to better capture severity-related patterns \cite{joshy2023dysarthria}. Additionally, Recurrent Neural Networks (RNNs) and Long Short-Term Memory (LSTM) networks have been utilized to model temporal dependencies, further enhancing classification performance \cite{bhat2020automatic, al2024detection, joshy2021automated}.

More recently, self-supervised learning techniques like wav2vec 2.0 \cite{baevski2020wav2vec} and HuBERT \cite{hsu2021hubert} have demonstrated their potential in pretraining models on large-scale speech corpora, improving feature extraction and transfer learning. Additionally, attention-based architectures and transformers \cite{shahamiri2023dysarthric} have enhanced long-range dependency modeling, further improving classification performance. A key challenge remains in developing speaker-independent (SI) models that generalize across diverse speech characteristics. To mitigate speaker variability and enhance model robustness, researchers have explored domain adaptation \cite{woszczyk20_interspeech}, multi-task learning \cite{xiong2024improving}, and adversarial training \cite{woszczyk20_interspeech}. However, data scarcity, severity class imbalance, and generalization to unseen speakers continue to pose significant challenges, necessitating further advancements in transfer learning and cross-corpus learning strategies.

In this study, we introduce DSSCNet (Dysarthric Speech Severity Classification Network), a novel deep learning architecture designed for speaker-independent (SI) dysarthric speech severity classification. The proposed model is specifically developed to address speaker variability, dataset biases, and class imbalance issues that often arise in dysarthria assessment. Unlike conventional approaches that rely on single-dataset training, DSSCNet is designed within a multi-corpus learning framework, leveraging data from multiple dysarthric speech datasets to enhance generalization across unseen speakers and severity levels.

\section{Methodology}
\label{sec:methodology}
\subsection{Proposed Classification Network: DSSCNet}
\label{subsec:DSSCnet}
\begin{figure*}[!ht]
    \centering
    \includegraphics[width=\linewidth]{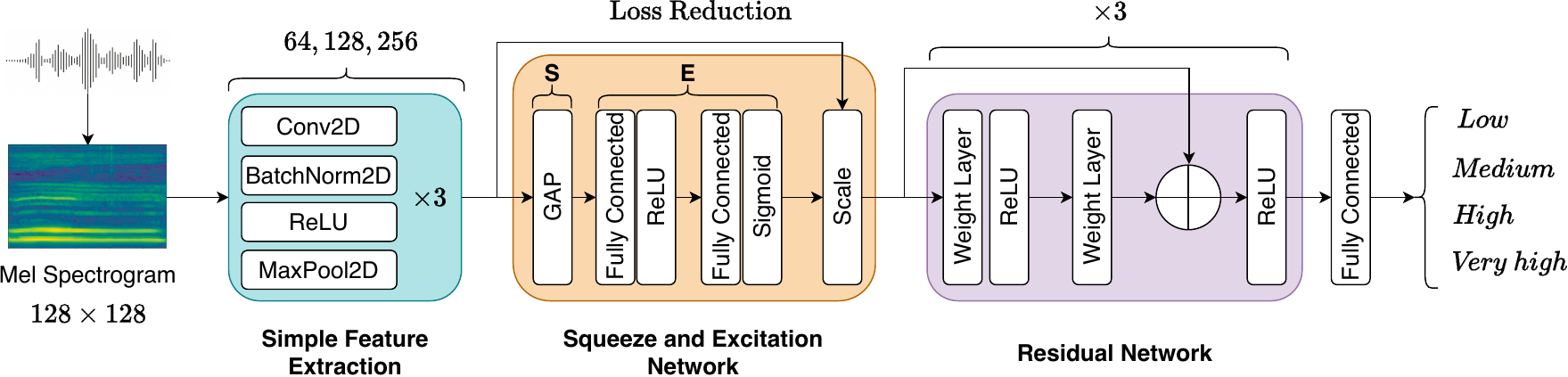}
    \caption{Overview of DSSCNet for dysarthria speech severity classification.}
    \label{fig:model-arch}
\end{figure*}

Dysarthric Speech Severity Classification Network, or in short DSSCNet, has an extensive architecture consisting of three parts: a simple feature extraction from the CNN backbone, the Squeeze and Excitation Network (SENet), and the Residual Network for complex feature extraction. These blocks help minimize losses during training and are capable of learning complex features, resulting in a model that accurately classifies severity. The overall architecture of DSSCNet is illustrated in Fig. \ref{fig:model-arch}.

\subsubsection{Simple Feature Extraction}
\label{subsubsec:simple-feature}
A speech signal is converted into a Mel spectrogram of size $128 \times 128$, serving as input to a CNN backbone for feature extraction. The CNN applies convolutional layers with Batch Normalization, progressively increasing feature channels from 64 to 256, capturing discriminative spectral and temporal patterns for severity classification. The extracted feature maps $f_0 \in \mathbb{R}^{H \times W \times C}$ encode refined speech representations, where $H$ and $W$ represent transformed time and frequency dimensions, and $C$ corresponds to the number of learned feature channels. These representations are further processed to improve classification performance.

\subsubsection{Loss Reduction using SENet}
\label{subsubsec:loss-reduction}
Squeeze and Excitation Networks (SENet) \cite{hu2018squeeze} enhance CNNs by modeling channel-wise inter-dependencies, allowing the network to emphasize key features. The SE block, its fundamental unit, consists of Squeeze and Excitation stages. Global Average Pooling (GAP) in the Squeeze stage reduces spatial dimensions, summarizing feature distributions per channel. The Excitation stage applies a gating mechanism to generate attention weights, determining channel importance. This GAP-based aggregation combined with a Sigmoid-activated gating mechanism dynamically refines feature prioritization, improving network performance.

The combination of the CNN-based feature extractor with the SENet forms our baseline model, referred to as CNN + SE, which serves as a comparative reference against the proposed DSSCNet architecture throughout our experiments.

\subsubsection{Residual Feature Extraction}
\label{subsubsec:residual-features}
Residual Networks \cite{he2016deep} utilize Residual Blocks with weighted layers, ReLU activation, and skip connections to learn residual functions, improving training efficiency. Instead of direct input-output mapping, residual functions capture differences, aiding gradient flow and mitigating vanishing gradients, allowing deeper architectures. Skip connections preserve the original input signal, enhancing convergence and generalization. Additionally, Adaptive Average Pooling (AAP) \cite{liu2018path} standardizes output dimensions, ensuring compatibility across datasets. The final stage outputs a probability distribution over severity levels, enabling accurate classification.

\subsection{Framework for Cross-corpus fine-tuning}
\label{subsec:proposed-framework}
\begin{figure}[!ht]
    \centering
    \includegraphics[width=\linewidth]{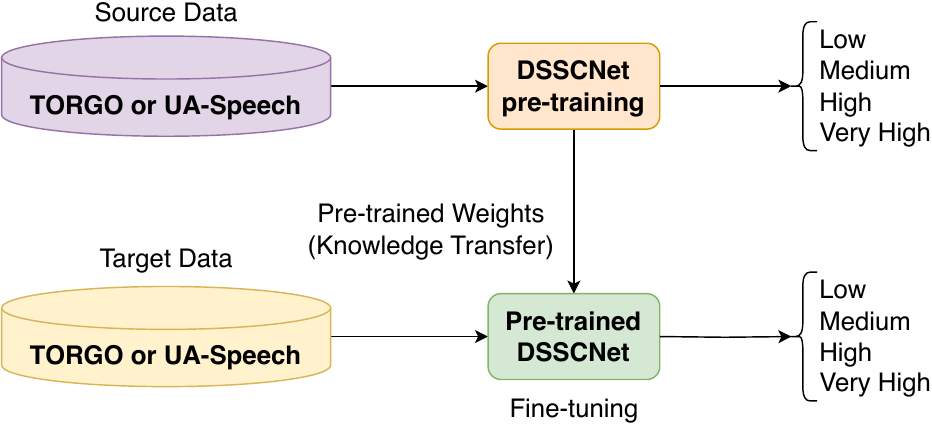}
    \caption{Block diagram of the DSSCNet fine-tuning framework for dysarthric speech severity classification.}
    \label{fig:framework}
    \vspace{-1em}
\end{figure}

Building upon the foundational concepts of cross-corpus transfer learning explored in prior studies on dysarthria detection \cite{wang2021unsupervised, wav2vecfinetune}, we propose a novel adaptation of this paradigm for the task of dysarthric speech severity classification. The framework, illustrated in Fig. \ref{fig:framework}, leverages pre-training and fine-tuning to improve the generalization and robustness of the proposed DSSCNet model across diverse speech corpora.

\section{Experimental Setup}
\label{sec:experimental-setup}

\subsection{Dataset Description}
\label{subsec:dataset}
This study utilizes the TORGO \cite{rudzicz2012torgo} and UA-Speech \cite{kim2008dysarthric} datasets, two publicly available dysarthric speech corpora containing recordings of individuals with varying levels of speech impairment. The speaker-wise severity levels and utterance distribution used in this study are detailed in Table \ref{tab:dataset}.

The \textbf{TORGO} dataset consists of recordings from 8 dysarthric speakers (3 female, 5 male) exhibiting low, medium, and high severity levels, along with 7 control speakers. The dataset includes a diverse range of speech samples, such as isolated words, sentences, and sustained vowels, recorded using both head-mounted and directional microphones at a 16 kHz sampling rate. 

The \textbf{UA-Speech} dataset comprises recordings from 15 dysarthric speakers (4 female, 11 male), providing a larger and more diverse corpus for dysarthric speech assessment. Speech samples include isolated words and short phrases, recorded at a 16 kHz sampling rate. Out of the total 15 dysarthric speakers, 12 speakers were selected for equal distribution of severity per speaker.

For SI evaluation in the both the datasets, one speaker per severity is taken in the test set, while the rest are used in the train set. For TORGO: 5 speakers for training and 3 for testing and for UA-Speech: 8 speakers for training and 4 for testing. This combination is repeated to ensure that each speaker was used for evaluation at-least once. A total of 18 and 81 unique combinations are created for TORGO and UA-Speech respectively.

\renewcommand{\arraystretch}{1.1}
\begin{table*}[!ht]
    \centering
    \caption{Speaker-wise utterance count and severity level description for both TORGO and UA-Speech dataset.}
    \scalebox{0.97}{
    \begin{tabular}{c|c|ccc|ccc|ccc|ccc}
        \hline

        \hline
        \multicolumn{2}{c|}{\textbf{Severity}} & \multicolumn{3}{c|}{Low} & \multicolumn{3}{c|}{Medium} & \multicolumn{3}{c|}{High} & \multicolumn{3}{c}{Very High} \\
        \hline

        \multirow{2}{*}{\textbf{TORGO}} & \textbf{Speaker} & F03 & F04 & M03 & F01 & M05 & - & M01 & M02 & M04 & - & - & - \\
        & \textbf{No. of Utterances} & 1075 & 667 & 800 & 228 & 573 & - & 739 & 766 & 652 & - & - & - \\
        \hline

        \multirow{2}{*}{\textbf{UA-Speech}} & \textbf{Speaker} & F05 & M08 & M09 & F04 & M05 & M11 & F02 & M07 & M16 & F03 & M04 & M01 \\
        & \textbf{No. of Utterances} & 5355 & 5355 & 5354 & 5251 & 5354 & 4590 & 5354 & 5354 & 4590 & 5182 & 3825 & 2805 \\
        \hline

        \hline
    \end{tabular}}
    \label{tab:dataset}
    \vspace{-1em}
\end{table*}

\subsection{Data Preprocessing}
\label{subsec:data-preprocessing}
A structured data preprocessing pipeline is applied for consistent feature extraction.
The complete duration of each waveform is inputted without any truncation or padding. A Mel spectrogram is then extracted using STFT with an FFT size of 256 and a hop length of 64 ms, utilizing 128 Mel filter banks. The log-scaled spectrograms are resized to a fixed $128 \times 128$ resolution using bilinear interpolation, maintaining consistency across samples. Finally, to match the DSSCNet input format, the spectrograms are expanded to 3 channels by replicating the single-channel Mel features across three dimensions. The DSSCNet model is trained using a batch size of 16, a learning rate of $1 \times 10^{-3}$, optimizer of Adam \cite{kingma2014adam}, and for 10 epochs. To optimize the performance of severity classification, the model is trained with the CrossEntropy loss function \cite{zhang2018generalized}, with class-specific weighting applied to address class imbalance. The weights are computed based on the number of samples in each severity class, ensuring that underrepresented classes contribute proportionally to the loss and improving the model’s ability to learn from all severity levels.



\subsection{Transfer-learning with DSSCNet}
\label{subsec:pretraining}


To improve performance in SI configurations, we adopt a cross-corpus transfer-learning strategy as discussed in Section \ref{subsec:proposed-framework}. DSSCNet is first pre-trained on the complete training dataset of one corpus either TORGO or UA-Speech to learn generalized representations of dysarthric speech across diverse severity levels and speaker characteristics. The pre-train model is then fine-tuned and evaluated on the SI sets of the other corpus, enabling us to assess the model's ability to generalize to unseen speakers under different recording and linguistic conditions.

To ensure clarity and reproducibility, the following two cross-corpus configurations are evaluated:

\subsubsection{UA-Speech $\rightarrow$ TORGO}
\label{subsubsec:pretrained-uaspeech}
DSSCNet is pre-trained on the full UA-Speech dataset and subsequently fine-tuned and evaluated on the SI sets of TORGO. This configuration assesses the model’s ability to adapt to a smaller corpus with distinct articulation patterns and recording environments.

\subsubsection{TORGO $\rightarrow$ UA-Speech}
\label{subsubsec:pretrained-torgo}
In the reverse configuration, DSSCNet is pre-trained on the TORGO dataset and then fine-tuned on the SI sets of UA-Speech. This setting presents a more challenging generalization task due to the larger speaker pool and greater variability in severity levels within UA-Speech.

To assess the effectiveness of pre-training, DSSCNet was tested on both TORGO and UA-Speech, achieving test accuracies of 97.66\% and 98.94\%, respectively. For both speaker-dependent and speaker-independent evaluations, 10\% of the training set was used as a validation set. The classification performance across severity levels is reported in Table \ref{tab:ua-speech_sd}.

\vspace{-3pt}
\renewcommand{\tabcolsep}{9pt}
\begin{table}[!ht]
    \centering
    \caption{Performance of DSSCNet on speaker dependent dysarthria speech severity classification on both TORGO and UA-Speech dataset.}
    \begin{tabular}{c|c|c}
    \hline

    \hline
         \multirow{2}{*}{\textbf{Severity}} & Accuracy (\%) & Accuracy (\%) \\
         & on TORGO & on UA-Speech \\
         \hline
         Low & 97.00 & 99.00 \\
         Medium & 99.00 & 99.00 \\
         High & 98.00 & 99.00 \\
         Very High & - & 98.00 \\
         \hline
         \textbf{Overall Accuracy} & \textbf{97.66} & \textbf{98.94} \\
         \hline

         \hline
    \end{tabular}
    \label{tab:ua-speech_sd}
\end{table}

\section{Results and Discussion}
\label{sec:results}
\begin{figure*}[!ht]
    \centering
    \subfigure[\label{subfig:torgo-baseline}]{\includegraphics[width=0.245\linewidth]{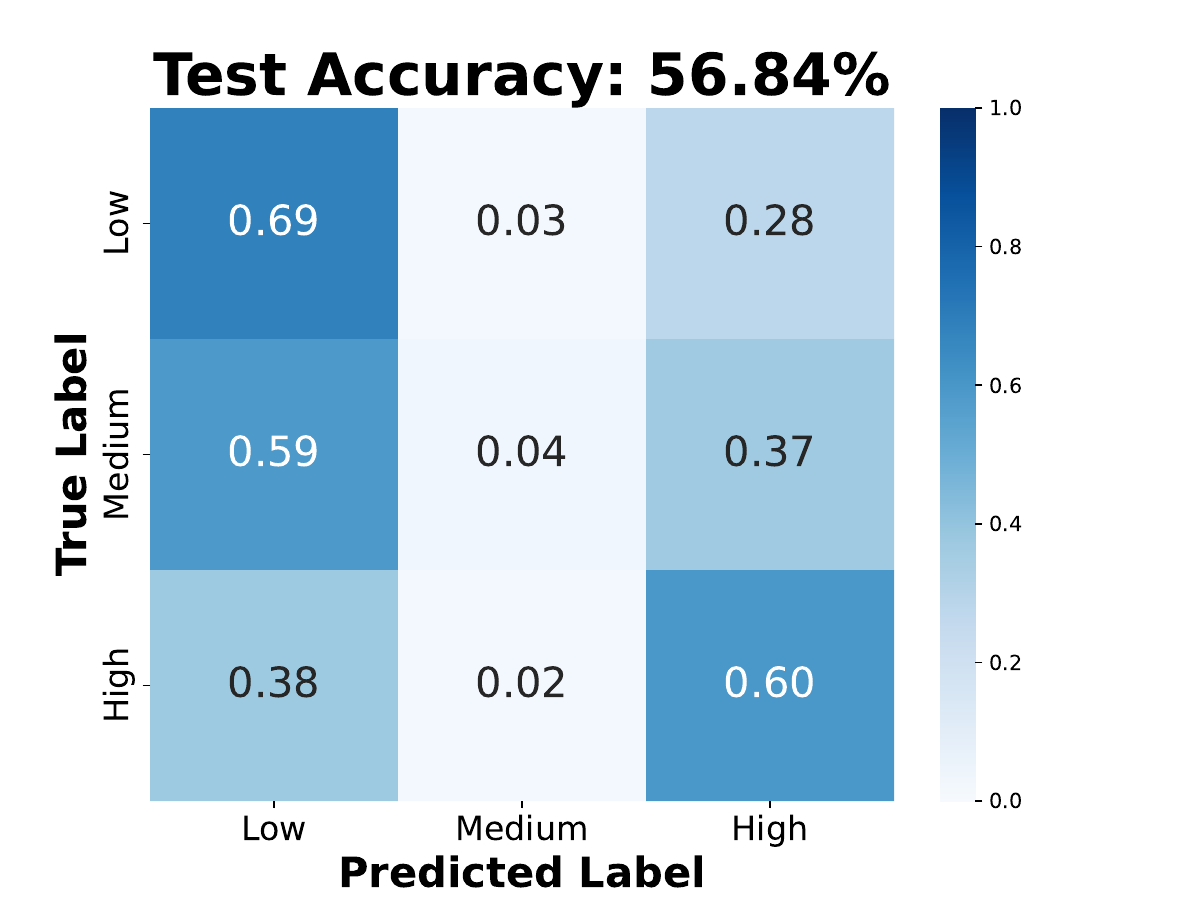}} \hfill
    \subfigure[\label{subfig:torgo-finetune}]{\includegraphics[width=0.245\linewidth]{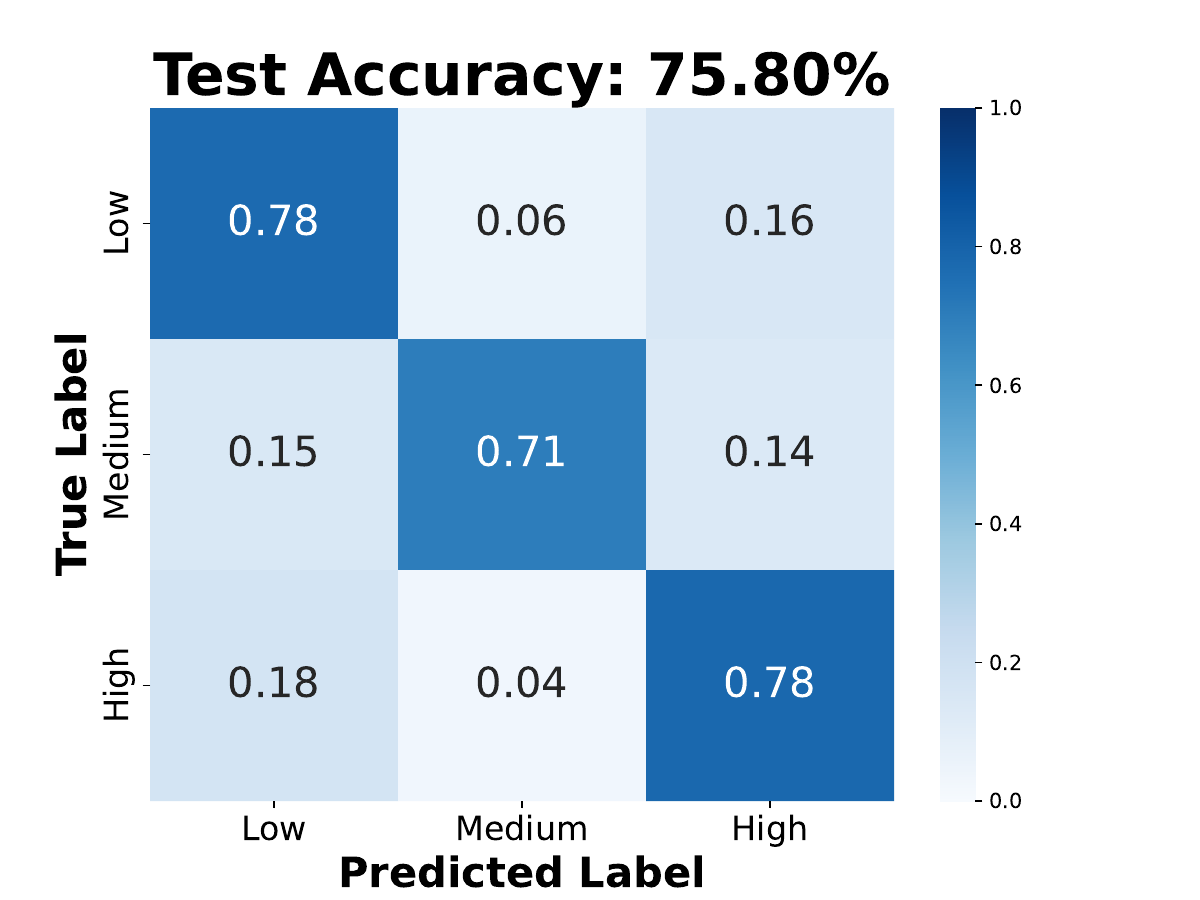}} \hfill
    \subfigure[\label{subfig:uaspeech-baseline}]{\includegraphics[width=0.245\linewidth]{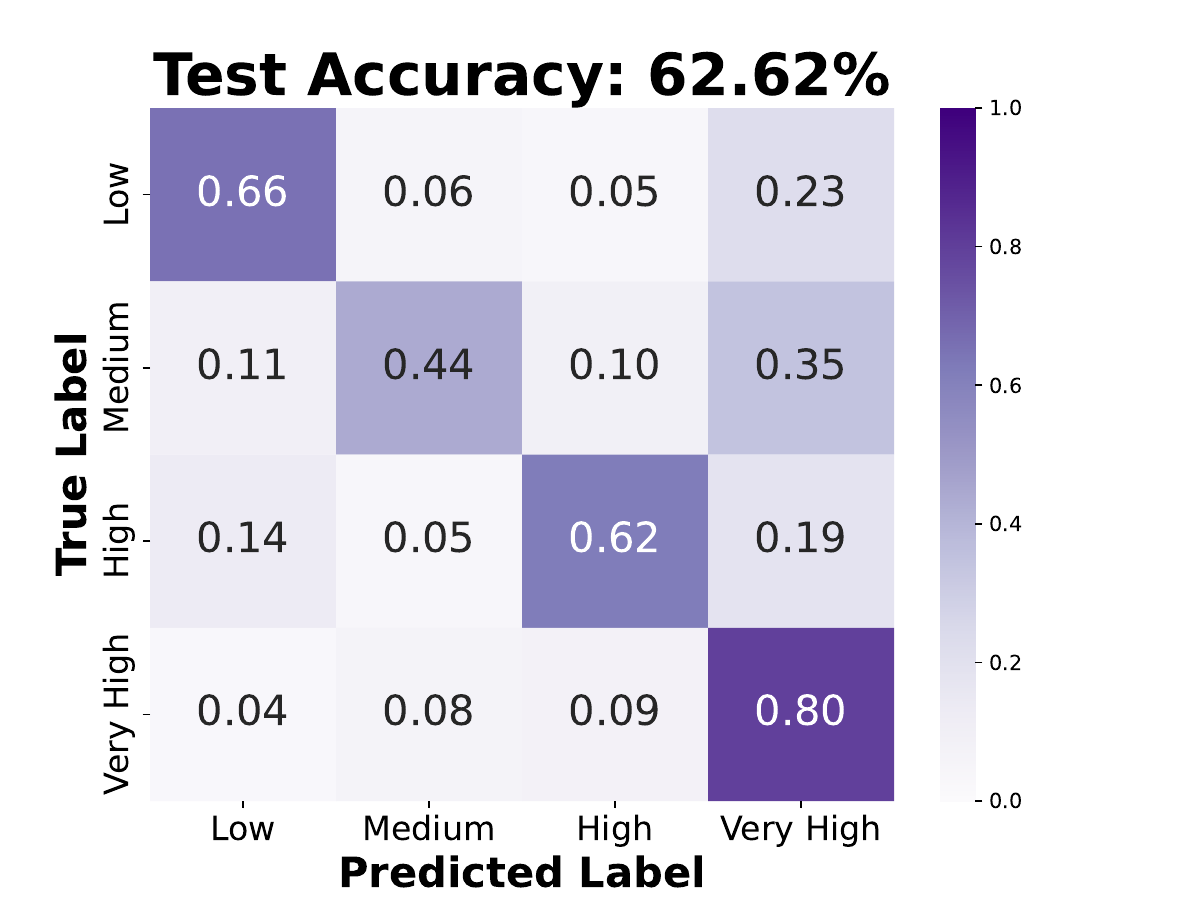}} \hfill
    \subfigure[\label{subfig:uaspeech-finetune}]{\includegraphics[width=0.245\linewidth]{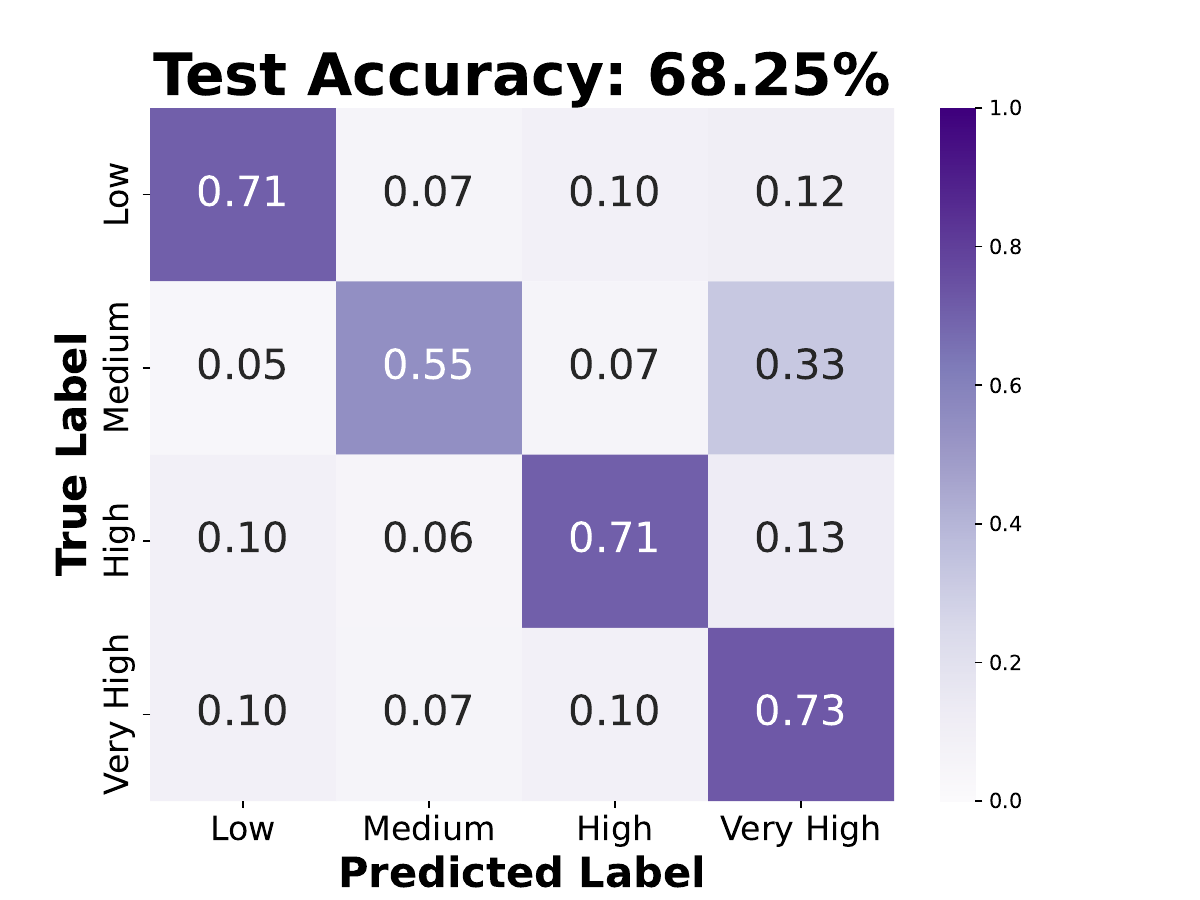}} \hfill
    \caption{Normalized confusion matrices of DSSCNet on TORGO: (a) w/o fine-tuning, (b) w/ fine-tuning averaged across 18 sets and UA-Speech: (c) w/o fine-tuning, (d) w/ fine-tuning averaged across 81 sets.}
    \label{fig:confusion-matrices}
    \vspace{-1em}
\end{figure*}

This section presents the experimental results evaluating DSSCNet for dysarthric speech severity classification. The model’s performance is assessed in terms of classification accuracy, generalization across unseen speakers, and the effectiveness of transfer learning strategies. Table \ref{tab:sota-comparison} summarizes the comparative classification accuracies across different architectures, while Figure \ref{fig:confusion-matrices} illustrates DSSCNet’s performance under various training configurations, such as fine-tuning and cross-corpus adaptation.

\subsection{Comparison with previous work}
\label{subsec:comparison-sota}
The performance of DSSCNet is evaluated against state-of-the-art architectures to assess its effectiveness in dysarthric speech severity classification. The comparison highlights DSSCNet’s ability to capture severity-related speech patterns, leveraging transfer learning and cross-corpus adaptation for improved SI performance.

\vspace{-3pt}
\begin{table}[!ht]
    \centering
    \caption{Speaker-Independent Classification Accuracy Comparison of DSSCNet with State-of-the-Art Methods on TORGO and UA-Speech.}
    \scalebox{0.94}{
    \begin{tabular}{c|c|c}
        \hline

        \hline
        \multirow{2}{*}{\textbf{Methods}} & \multicolumn{2}{c}{Accuracy (\%)} \\ \cline{2-3}
        & \textbf{TORGO} & \textbf{UA-Speech} \\
        \hline
        CNN (Mel-Spectrogram) \cite{8884185} & 49.27 & - \\
        CNN with HuBERT \cite{javanmardi2024pre} & 49.83 & 48.01 \\
        DNN with MFCC \cite{joshy2022automated} & - & 49.22 \\
        CNN with DeepSpeech \cite{9054492} & - & 53.90 \\
        \hline
        \hline
        DSSCNet (Proposed) & \textbf{56.84} & \textbf{62.62} \\
        DSSCNet (Proposed) + fine-tuning & \textbf{75.80} & \textbf{68.25} \\
         \hline

         \hline
    \end{tabular}}
    \label{tab:sota-comparison}
\end{table}

Table \ref{tab:sota-comparison} compares DSSCNet with existing approaches under SI settings. On TORGO, DSSCNet achieves an accuracy of 56.84\%, providing an absolute improvement of 7.57\% over CNN (Mel-Spectrogram) \cite{8884185}, which attains 49.27\%, and a 7.01\% improvement over CNN with HuBERT \cite{javanmardi2024pre}, which reaches 49.83\%. On UA-Speech, DSSCNet attains 62.62\%, marking a gain of 8.72\% over CNN with DeepSpeech \cite{9054492} at 53.90\% and 13.40\% over CNN with HuBERT \cite{joshy2022automated}. These findings underscore the effectiveness of DSSCNet’s architectural design in capturing severity-specific speech characteristics and generalizing across unseen speakers. 

After fine-tuning, DSSCNet achieves an accuracy of 75.80\%, providing a 26.53\% improvement over CNN (Mel-Spectrogram) \cite{8884185}. DSSCNet also outperforms CNN with HuBERT \cite{javanmardi2024pre}, which achieves 49.83\%, by 25.97\%. On the UA-Speech dataset, DSSCNet reaches 68.25\%, surpassing CNN with DeepSpeech \cite{9054492} by 14.35\% (from 53.90\%) and outperforming CNN with HuBERT by 20.24\%. It also exceeds the performance of a DNN model utilizing MFCC-based i-vectors \cite{joshy2022automated}, which records an accuracy of 49.22\%, by 19.03\%. These results highlight the effectiveness of the proposed fine-tuning strategy in enhancing model generalization and improving severity classification across different dysarthric speech datasets.


\subsection{Ablation Study}
\label{tab:ablation-study}
To evaluate the impact of architectural enhancements on dysarthric speech severity classification, we compare DSSCNet against our baseline CNN + SE model. This network, illustrated in Section \ref{subsubsec:loss-reduction}, consists of a series of convolutional layers for hierarchical feature extraction, followed by SENet to refine channel-wise feature importance. The SE mechanism applies global average pooling to capture spatial dependencies, followed by a gating function to reweigh feature channels dynamically. This allows the network to focus on the most relevant spectral representations while improving feature discrimination for dysarthric speech.

We conduct experiments with and without fine-tuning utilizing the transfer learning framework depicted in Fig. \ref{fig:framework} to assess the effectiveness of transfer learning within our framework. The classification accuracies obtained using the CNN + SE model and DSSCNet are reported in Table \ref{tab:accuracy-table}. The results demonstrate that DSSCNet significantly outperforms the baseline, even w/o fine-tuning, highlighting the architectural feat of the proposed model.

\vspace{-3pt}
\renewcommand{\tabcolsep}{15pt}
\begin{table}[!ht]
    \centering
    \caption{Comparison of classification accuracies between DSSCNet and a simple CNN network with SE Block on TORGO dataset.}
    \begin{tabular}{c|c|c}
        \hline

        \hline
        \textbf{Method} & \textbf{Fine-tuned} & \textbf{Accuracy (\%)} \\
        \hline
         CNN + SE & \xmark & 44.04 \\
         CNN + SE & \cmark & 52.37 \\
         DSSCNet & \xmark & 56.84 \\
         DSSCNet & \cmark & \textbf{75.80} \\
         \hline

         \hline
    \end{tabular}
    \label{tab:accuracy-table}
\end{table}

\subsection{Effect of cross-corpus learning}
\label{subsec:discussion}
Figure \ref{fig:confusion-matrices} illustrates the impact of cross-corpus learning on dysarthric speech severity classification, with confusion matrices for TORGO (Fig. \ref{subfig:torgo-baseline}, Fig. \ref{subfig:torgo-finetune}) and UA-Speech (Fig. \ref{subfig:uaspeech-baseline}, Fig. \ref{subfig:uaspeech-finetune}). Fine-tuning significantly improves classification performance across both datasets, reducing misclassification rates and enhancing generalization. Comparing Fig. \ref{subfig:torgo-baseline} and Fig. \ref{subfig:torgo-finetune} on the TORGO dataset, it is evident that fine-tuning Fig. \ref{subfig:torgo-finetune} improves overall accuracy, particularly for the medium severity class, which previously had higher confusion with adjacent severity levels. Similarly, the comparison between Fig. \ref{subfig:uaspeech-baseline} and Fig. \ref{subfig:uaspeech-finetune} on UA-Speech demonstrates that cross-corpus fine-tuning enhances classification performance, especially for medium and high severity levels, where subtle variations in dysarthric speech are more challenging to distinguish.

Furthermore, models without fine-tuning (Fig. \ref{subfig:torgo-baseline}, Fig. \ref{subfig:uaspeech-baseline}) struggle more with adjacent severity levels, whereas fine-tuned models (Fig. \ref{subfig:torgo-finetune}, Fig. \ref{subfig:uaspeech-finetune}) show reduced misclassification, confirming that cross-corpus learning enhances robustness and speaker-independent generalization.

\section{Conclusion}
\label{sec:conclusion}
This study introduced DSSCNet, a deep learning model for SI dysarthric speech severity classification, leveraging multi-corpus and transfer learning to improve generalization. Experimental results showed that pre-training on one dataset and fine-tuning on another enhanced classification accuracy, reducing misclassification and improving robustness to speaker variability. The findings highlight the effectiveness of cross-corpus adaptation in addressing dataset limitations and class imbalance, making DSSCNet a strong candidate for automated dysarthria severity assessment. Future work will explore self-supervised learning and multi-modal approaches to further advance assistive speech technologies.

\bibliographystyle{IEEEbib}
\bibliography{refs}

\end{document}